# A New Multi-Level Hazy Image and Video Dataset for Benchmark of Dehazing Methods

*Bedrettin Çetinkaya[1], Yücel Çimtay[2], Fatma Nazli Günay[3], and Gökçe Nur Yılmaz[4]

[1] Computer Engineering, Ted University, Ankara, Turkey
`1*bedrettin.cetinkaya@tedu.edu.tr`
[2] Computer Engineering, Ted University, Ankara, Turkey
`2yucel.cimtay@tedu.edu.tr`
[3] Computer Engineering, Ted University, Ankara, Turkey
`3fnazli.gunay@tedu.edu.tr`
[4] Computer Engineering, Ted University, Ankara, Turkey
`4gokce.yilmaz@tedu.edu.tr`

* Corresponding Author

**Abstract.** The changing level of haze is one of the main factors which affects the success of the proposed dehazing methods. However, there is a lack of controlled multi-level hazy dataset in the literature. Therefore, in this study, a new multi-level hazy color image dataset is presented. Color video data is captured for two real scenes with a controlled level of haze. The distance of the scene objects from the camera, haze level, and ground truth (clear image) are available so that different dehazing methods and models can be benchmarked. In this study, the dehazing performance of five different dehazing methods/models is compared on the dataset based on SSIM, PSNR, VSI and DISTS image quality metrics. Results show that traditional methods can generalize the dehazing problem better than many deep learning based methods. The performance of deep models depends mostly on the scene and is generally poor on cross-dataset dehazing.

## 1 Introduction

Different conditions that affect the scattering of the light such as haze, fog, and smoke can significantly reduce information contained in the captured scene. This loss of information causes poor performance in many computer vision problems such as object detection/tracking/classification [1, 2, 3], depth estimation [4], remote sensing [5] and etc. Therefore, reconstructing hazy-free data have the potential to improve the performance of other computer vision problems.

Traditional approaches try to generate hazy-free images using hand-crafted features/priors, such as Dark Channel Prior (DCP) [6], Color Attenuation Prior (CAP) [7], color- lines [8] and etc. On the other hand, the modern methods [9, 10, 11, 12, 13], are based on deep learning and they re- quire both hazy-free and hazy data for training. Therefore, the datasets have a big impact on the model performance. On the other hand, evaluating the cross-dataset performance of models gives information about their generalization capabilities. To this end, we published a new foggy image dataset, Indoor-Multi-Level Fog Dataset (IMFD), for image dehazing problem and compare several state of the art methods on IMFD.

## 2 Related Work

Image dehazing is one of the most attractive problems in computer vision. With the rise of deep learning, neural networks-based solutions have become more dominant rather than traditional methods such as DCP [6], CAP [7] and etc. There- fore, we divided previous studies into two main groups: (i) Traditional and (ii) Deep Learning-based methods.

### 2.1 Traditional Methods

Most of the traditional methods rely on prior information such as image saturation/attenuation, pixel values in the RGB channels and etc. and they try to model haze under the assumption of homogeneous distribution of haze, unlike the real-world scenario. He et al. [6] proposed one of the most important traditional methods named Dark Channel Prior (DCP). This method dehazes hazy images estimating transmission maps and atmospheric light. Due to the simplicity and efficiency of DCP algorithm, there are many improved versions [14, 15, 16, 17, 18]. Also, Zhe et al. [7] use Color Attenuation Prior (CAP). They use CAP to estimate the depth map of a hazy image and hazes are removed using this depth information. Moreover, Fattal et al. [8] exploit color-lines that come from 1D distribution of RGB pixel values. This color-line can successfully estimate transmission offsets of hazy images. Also, Berman et al. [19] remove haze using haze-lines which enables estimating the distance map and the atmospheric light.

### 2.2 Deep Learning-Based Methods

Deep learning-based techniques have gained increasing popularity in recent times due to their successful performance. Cai et al. [9] introduced a CNN, DehazeNet, that estimates a medium transmission map to reconstruct haze-free images. Also, Ullah et al. propose a lightweight version of DehazeNet. Unlike these works, Li et al. [20] directly estimated haze-free images without finding the output of any intermediate step such as transmission/distance map. Zhang et al. [10] proposed a GAN to jointly learn the atmosphere light and the transmission map. Singh et al. [21] employed GAN-based architecture which aims to learn multi-level complexity and multi-scale spatial features. Also, Wu et al. [11] successfully applied contrastive learning for image dehazing problem. Jin et al. [22] estimated uncertainty maps that are used as attention coefficients for the density of haze. Tran et al. [23] proposed a GAN model with DCP. Song et al. [12] used a vision transformer-based architecture to remove haze. Jiang et al [13] proposed a hybrid model which aims to de- hazes hazy images and refines the outputs of dehazing step.

## 3     Datasets

This section reviews popular datasets used in the image dehazing problem.

**Chic:** Color Hazy Image for Comparison Dataset, CHIC, [24] provides fog-free RGB images and their hazed versions with nine different haze levels in a well-controlled indoor environment. Also, it contains three different scenes and two different camera positions which are static and dynamic. 9 different images with different haze levels are presented. Scenes are cropped where full images size is 6000 x 4000. Scene-1_cropped images size is 1566 x 2244 and Scene-2 is 2010 x 3528.

**Dense-Haze:** Dense-Haze Dataset [25] contains 33 paired hazy and fog-free ground-truth images. It contains both indoor and outdoor scenes. Hazy images were created using professional haze machines and they have homogeneous haze distributions. The foggy sceneries were captured by using actual haze that was produced by expert haze generators. The identical visual information was recorded under the same lighting conditions in the respective haze-free and haze-filled scenarios. The goal of the Dense-Haze dataset is to greatly advance single-image dehazing state-of-the-art by pushing robust approaches for realistic and diverse hazy scenarios.

**D-Hazy:** D-Hazy Dataset [26] contains 1409 hazy and corresponding fog-free images. It used NYU Depth [27] and Middelbury [28] datasets because hazy images were generated synthetically using depth information in these datasets. Only indoor scene is available in this dataset. Due to the fact that in a hazy media the scene brightness attenuates with distance, researchers are able to produce a comparable hazy scene with high accuracy based on the depth information and utilizing the physical model of a hazy medium.

**I-Haze:** I-Haze Dataset [29] contains 35 paired hazy and haze-free ground-truth images. All scene contains a *MacBeth* color checker to simplify the color calibration process. Also, the same illumination conditions were used for all data.

**NH-Haze:** NH-Haze Dataset [30] contains 55 paired hazy and fog-free ground-truth images and these hazy images are created using a professional haze generator. Moreover, hazy images reflect real-world scenarios because these images have non-homogeneous haze distributions. Also, real hazy data without ground-truths are available for the evaluation process. All images were collected from the outdoor environment.

**O-Haze:** O-Haze Dataset [31] contains 45 paired hazy and fog-free ground-truth images from the outdoor environment. Also, it contains different illumination conditions because data were collected under different sunlight levels. Using traditional image quality measurements like PSNR, SSIM, and CIEDE2000, O-HAZE is used to compare a sample collection of cutting-edge dehazing approaches.

**Revide:** Unlike the previous image dehazing datasets, Revide Dataset [32] contains paired videos with hazy and haze-free data. Data were collected using 47 different scenes and each scene contains different haze levels and illumination conditions. By capturing the same scene (with or without haze) twice using a well-designed video acquisition system, they collect matched real-world hazy and haze-free movies that are precisely aligned. They also create a Confidence Guided and Improved Deformable Network (CG-IDN) for video dehazing in order to address the difficulty of utilizing temporal redundancy among the hazy frames.

## 4    IMF Dataset (IMFD)

Previously stated datasets are mostly based on static images and there is a lack of foggy video data. Another shortcoming is that they do not provide foggy scenes with a controlled level of fog except [24, 33]. To overcome that issues and enrich the foggy dataset literature, in this study, we present a new indoor foggy imagery dataset (Indoor-Multi-Level Fog Dataset) which was created in a light and fog-controlled environment. We present two different scenes which include objects with various distances from the camera. In addition, we provide the ground truth as both image and video data.

Hanwha Wisenet QNO-7082R Network camera [34] was used to capture the imagery of the scenes. Before recording, the camera was focused on the scenes. Video resolution was set as 1080p (1920 x 1080) with a ratio of 16:9. Frame rate and video coding standard were set as 60 fps and H.264 respectively. IR filter was turned off. The camera was connected remotely and during recording, all the conditions including the position of the camera, were the same except the level of the fog. First of all, the ground truth scene was captured and then the room was filled with fog up to the scene that cannot be seen by the eye. Then recording was restarted and maintained up to the fog density reduces and the scene approaches the ground truth.

Quenlite QL-2000RGB Fog Machine [35] was used for the purpose of fog production. This machine can be con- trolled remotely. Its fog emittance capacity and spraying distance are 560 m3 / min and 8m, respectively. In one cycle, it can continuously emit for 50 seconds. Then it cools down and is ready to be triggered again. We repeated the cycle up to the fog becomes an opaque layer. We recorded the foggy scene during the fog was being evacuated from the room.

We captured the imagery in a rectangular room that is isolated from the environment so that no daylight comes in. The lamps on the ceiling were used to enlighten the room. In this way, the amount of light in the room was kept stable. We created 2 different scenes with different levels of fog. The ground truth images and the depth of the objects are shown in Figure 1 for scenes 1 and 2. We positioned different objects at different depths which is a crucial concept to measure the success of any dehazing method in terms of improving the visual quality and increasing the visibility distance.

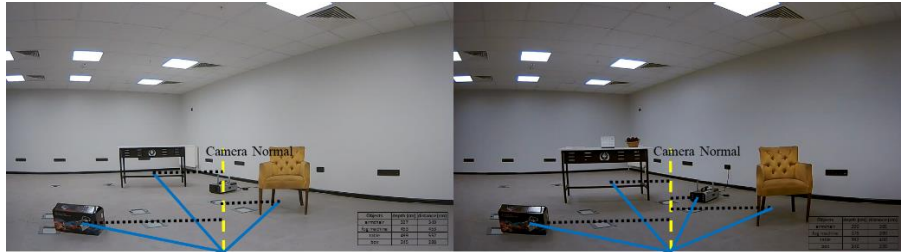

**Fig. 1.** Scenes 1 and 2, and the depth of the objects with respect to the camera, respectively.

The imagery of the scenes was captured while the fog was evacuating the room. The evacuation of the fog is controlled by a ventilation system. Therefore evacuation speed was constant with respect to time. After saving the recording we split the total video duration to 10 equal-length windows and snap a video frame corresponding to the ends of each interval. Figure 2 shows the frame selection process where fr is the frame rate of the video and L is the duration of the video. There- fore, the frame number (FN) corresponding to each fog level can be found by using Equation 1 where i and NL hold the fog level and the number of fog levels respectively.

$$FN_i = (NL - 1) * fr * L, i = 1, 2, 3, \ldots, NL \qquad 1$$

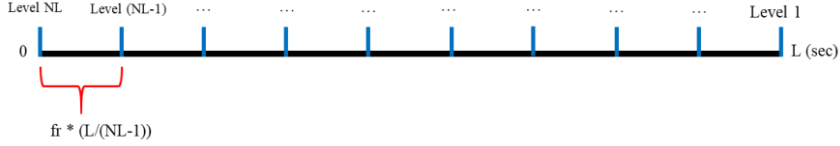

**Fig. 2.** Frame rate selection process.

Finally, we classified the video into 10 fog levels which are shown in Figure 3. Therefore for each scene, we provide 10-level foggy images (with ground truth) and also the labeled foggy videos with respect to the starting and ending times of different fog levels.

## 5       Experiment

We evaluate our dataset using state-of-the-art methods published with official pre-trained weights for deep learning- based models. These methods are: EDN-GTM [23], AECR- Net [11], Light-DehazeNet [36], DehazeFormer [12]. Moreover, we used different pre-trained models of each of these studies to see their generalization performance. We also evaluate the results of DCP [6] and $\omega$ is set to 0.45. DCP is one of the fundamental dehazing methods which uses atmospheric light scattering model. Air light is estimated by using the dark channel pixel information. We use Peak signal to noise ratio (PSNR), Structural similarity index measure (SSIM), Visual saliency-induced index (VIS) [37] and Deep Image Structure and Texture Similarity (DISTS) [38] metrics for the evaluation process as shown in Table 1. Peak signal-to-noise ratio (PSNR) is an expression for the ratio of the greatest potential value (power) of a signal to the power of the noise that distorts it and lowers the quality of its representation. The SSIM index is used to calculate how similar two images are to one another. A starting uncompressed or distortion-free image is used as the basis for the SSIM's prediction of image quality.

According to Table 1, EDN-GTM [23] trained with OHaze dataset performs better than all other methods for scene 1 in general. For scene 2, DehazeFormer [12] and EDN-GTM [23] have the best performances. On the hand, DCP [6] has considerable performance for both scenes 1 and 2. Other deep learning-based methods except for EDN-GTM and DehazeFormer have low generalization performance.

Moreover, we provide visual results for each method presented in Table 1, as shown in Figure 4. We consider only the best results for each method among the pre-trained weights.

## 6      Results and Discussion

Level of the haze is one of the most important effect which determines the success of any dehazing method.

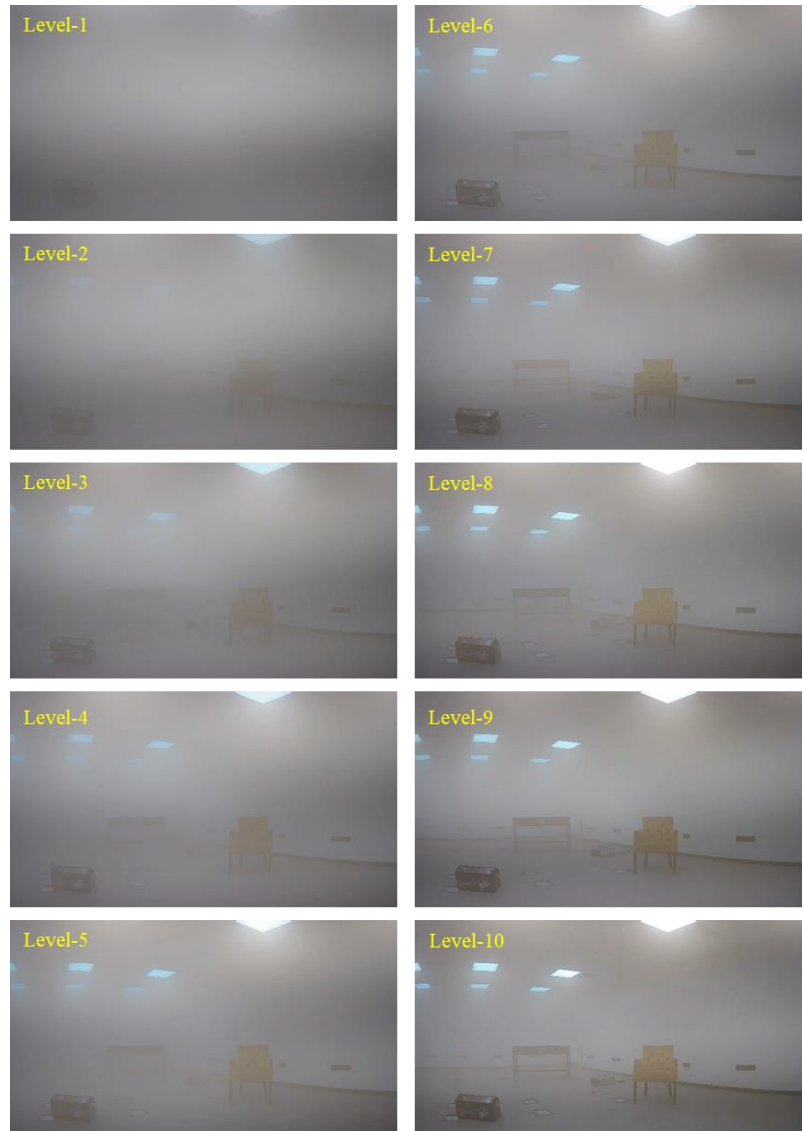

**Fig. 3.** Fog Levels from 1 to 10 for Scene 1.

Most of the methods perform well when haze level is low. However when the level is increased, the success drops dramatically. Therefore, in this study, we present a new multi-level color hazy image dataset which includes two scenes with different object distances from the camera. 5 different methods are compared on the dataset and results are presented. Performance of the dehazing methods can be represented by their generalization capability and adaptation to new hazy scenes. Compared to traditional methods, deep models perform better when tested on the same dataset they are trained with. However, cross- dataset dehazing performance is generally lower than the traditional methods. In the future, more effort should be given to develop deep models with high capacity of generalization.

**Fig. 4.** Visual results for state-of-the-art methods. We visualize only the best result presented in Table 1 for each method.

Table 1 Results for state-of-the-art methods in terms of SSIM, PSNR, VIS and DISTS metrics for both ITUFD scenes 1 and 2. While bold numbers show the best results for the corresponding level, metric, and scene, red numbers represent the second-best results. Also, ∗ means that the corresponding dataset was collected with public hazy images from internet resources.

| Method | Training Datasets | ITUFD Scene | Level 2 | | | | Level 5 | | | | Level 8 | | | |
|---|---|---|---|---|---|---|---|---|---|---|---|---|---|---|
| | | | SSIM | PSNR | VSI | DISTS | SSIM | PSNR | VSI | DISTS | SSIM | PSNR | VSI | DISTS |
| DCP (CVPR 2009) [6] | - | 1 | **0.85** | 16.35 | 0.90 | 0.58 | 0.83 | 15.10 | 0.90 | 0.46 | 0.84 | 15.47 | 0.90 | 0.40 |
| AECR-Net (CVPR 2021) [11] | DenseHaze | 1 | 0.42 | 12.08 | 0.83 | 0.37 | 0.49 | 11.61 | 0.85 | 0.37 | 0.50 | 12.22 | 0.84 | 0.35 |
| | ITS | 1 | 0.74 | 14.25 | 0.89 | 0.50 | 0.74 | 14.16 | 0.90 | 0.42 | 0.71 | 12.93 | 0.91 | 0.36 |
| | NHHaze | 1 | 0.44 | 10.15 | 0.81 | 0.39 | 0.43 | 9.32 | 0.86 | 0.37 | 0.46 | 9.89 | 0.88 | 0.37 |
| Light-D.Net (IEEE TIP 2021) [36] | * | 1 | 0.49 | 11.15 | 0.88 | 0.48 | 0.51 | 11.41 | 0.88 | 0.42 | 0.48 | 11.24 | 0.89 | 0.40 |
| EDN-GTM (ISCSI 2022) [23] | DenseHaze | 1 | 0.74 | 13.09 | 0.89 | 0.36 | 0.78 | 13.43 | **0.91** | **0.33** | 0.75 | 12.99 | 0.91 | **0.31** |
| | IHaze | 1 | 0.82 | 14.92 | 0.90 | 0.43 | 0.80 | 13.96 | **0.91** | 0.36 | 0.81 | 14.35 | **0.92** | 0.32 |
| | NHHaze | 1 | 0.80 | 13.76 | 0.89 | 0.44 | **0.84** | 14.13 | 0.90 | 0.41 | **0.85** | 14.56 | 0.91 | 0.38 |
| | OHaze | 1 | **0.85** | **16.38** | 0.88 | 0.52 | **0.84** | **15.49** | 0.90 | 0.43 | **0.85** | **16.20** | 0.91 | 0.39 |
| DehazeFormer (IEEE TIP 2023) [12] | Indoor | 1 | 0.73 | 12.36 | 0.89 | 0.50 | 0.65 | 10.50 | 0.90 | 0.42 | 0.70 | 11.26 | 0.91 | 0.39 |
| | Reside6k | 1 | 0.75 | 12.23 | **0.93** | **0.31** | 0.66 | 10.60 | 0.90 | 0.43 | 0.70 | 11.53 | 0.91 | 0.37 |
| | RSHaze | 1 | 0.83 | 15.06 | 0.86 | 0.51 | 0.82 | 14.92 | 0.86 | 0.41 | 0.83 | 15.22 | 0.87 | 0.39 |
| DCP (CVPR 2009) [6] | - | 2 | **0.83** | 17.39 | 0.89 | 0.60 | 0.81 | 16.72 | 0.92 | 0.41 | 0.84 | 17.93 | 0.93 | 0.32 |
| AECR-Net (CVPR 2021) [11] | DenseHaze | 2 | 0.42 | 13.50 | 0.84 | 0.40 | 0.48 | 12.34 | 0.84 | 0.36 | 0.56 | 15.19 | 0.89 | 0.29 |
| | ITS | 2 | 0.68 | 15.09 | 0.89 | 0.52 | 0.60 | 13.06 | 0.91 | 0.38 | 0.70 | 15.66 | 0.92 | 0.29 |
| | NHHaze | 2 | 0.33 | 11.64 | 0.81 | 0.41 | 0.46 | 10.01 | 0.85 | 0.35 | 0.51 | 11.56 | 0.88 | 0.31 |
| Light-D.Net (IEEE TIP 2021) [36] | * | 2 | 0.55 | 9.12 | 0.88 | 0.49 | 0.55 | 9.25 | 0.90 | 0.41 | 0.56 | 9.33 | 0.91 | 0.35 |
| EDN-GTM (ISCSI 2022) [23] | DenseHaze | 2 | 0.76 | 16.45 | 0.89 | 0.40 | 0.79 | 16.56 | **0.93** | **0.30** | 0.79 | 17.34 | 0.93 | 0.26 |
| | IHaze | 2 | 0.76 | 15.90 | 0.90 | 0.47 | **0.82** | 17.19 | **0.93** | 0.33 | 0.80 | 17.57 | **0.95** | **0.24** |
| | NHHaze | 2 | 0.70 | 14.11 | 0.86 | 0.45 | 0.79 | 14.53 | 0.91 | 0.39 | **0.85** | 18.18 | 0.94 | 0.29 |
| | OHaze | 2 | **0.83** | **17.52** | 0.89 | 0.54 | 0.82 | 16.62 | 0.91 | 0.41 | **0.85** | **19.05** | 0.92 | 0.32 |
| DehazeFormer (IEEE TIP 2023) [12] | Indoor | 2 | 0.74 | 15.07 | 0.89 | 0.54 | 0.74 | 14.76 | 0.92 | 0.36 | 0.81 | 17.08 | 0.94 | 0.28 |
| | Reside6k | 2 | 0.73 | 14.40 | **0.93** | **0.27** | 0.56 | 12.18 | 0.91 | 0.43 | 0.68 | 13.73 | 0.92 | 0.34 |
| | RSHaze | 2 | **0.83** | 17.02 | 0.88 | 0.54 | **0.82** | **17.38** | 0.89 | 0.39 | **0.85** | 18.91 | 0.91 | 0.30 |

## 7      Conclusion

In this study, a new multi-level hazy image and video dataset is proposed. By controlling the speed of the evacuation of the fog from the room, the fog level is adjusted. There are 10 level of foggy images of two scenes. DCP and the state-of-the art deep learning based dehazing methods are tested on this new dataset. According to the results, DCP and EDN-GTM achieves promising results. Another important result is that deep learning based methods have limited generalization capability when tested on a new dataset. This study is important and valuable since it brings a new multi-level hazy dataset to dehazing literature. The state of the art methods can be benchmarked on this dataset.